\documentclass[letterpaper,12pt]{article}
\usepackage{osajnl}
\usepackage[dvips,colorlinks=true,bookmarks=false,citecolor=blue,
urlcolor=blue,pdfstartview=FitH]{hyperref}

\begin{document}

\title{Negative Refraction Gives Rise to the Klein Paradox}

\author{Durdu \"O. G\"uney*}
\address{Department of Electrical and Computer Engineering, University
of California, San Diego, 9500 Gilman Dr., La Jolla, California, 92093-0409}
\address{Department of Mathematics, University
of California, San Diego, 9500 Gilman Dr., \\
La Jolla, California, 92093-0112}
\email{dguney@ucsd.edu}

\author{David A. Meyer}
\address{Department of Mathematics, University
of California, San Diego, 9500 Gilman Dr., \\
La Jolla, California, 92093-0112}
\email{dmeyer@math.ucsd.edu}

\maketitle

\newpage


\noindent {\bf Abstract} \\
\\
\noindent Electromagnetic negative refraction in metamaterials has attracted increasingly great interest, since its first experimental verification in 2001. It potentially leads to the applications superior to conventional devices including compact antennas for mobile stations, imaging beyond the diffraction limit, and high-resolution radars, not to mention the anamolous wave propagation in fundamental optics. Here, we report how metamaterials could be used to simulate the ``negative refraction of spin-zero particles interacting with a strong potential barrier", which gives rise to the Klein paradox$-$a counterintuitive relativistic process. We address the underlying physics of analogous wave propagation behaviours in those two entirely different domains of quantum and classical.

\section{Introduction}
About thirty years after Veselago's theoretical prediction$^{1}$, an effective left-handed material (LHM) was proposed, consisting of a periodic array of split ring resonators (SRRs) and continuous wires, which manifests refractive index $n<0$ for a certain frequency region$^{2-3}$. First, experimental verification of LHM was achieved in 2001 in San Diego$^{4}$. Subsequent efforts pushed the limits to telecom band and even to visible spectrum$^{5-9}$. In an entirely different domain, here we report how the LHMs could indeed simulate the negative refraction of spin-zero particles interacting with a strong potential barrier. Hence, they give rise to the Klein paradox$-$a counterintuitive phenomenon in relativistic quantum mechanics. We explain the underlying physics of the similar wave propagation behaviors in the Klein-Gordon and the Maxwellian pictures. Our work has implications to the analog quantum simulations of many exotic phenomena in quantum electrodynamics with relatively simple optical benchtop experiments and metamaterial designs on demand.

By taking the three spatial coordinates into account$^{13}$, the generation of scalar particles in the presence of a strong Klein step potential$^{10-12}$ has been considered. To the best of our knowledge, such a two-dimensional (2D) scattering from a semi-infinite Klein step potential has not been considered. Hence, the negative refraction is treated explicitly here first, inspired by the works about its optical counterpart, LHM$^{1,14-16}$. It is amazing that the 2D case not only verifies the
Klein paradox existing in one dimension, but it also shows the negative refraction of matter waves. Besides, in the territory of electronics, also inspired by LHMs, negative refraction was recently predicted theoretically in a $Science$ paper by Cheianov et al.$^{17}$ in a monolayer of graphite (graphene). Graphene not only leads to the focusing of electrons$^{17}$ similar to the perfect optical lens, but also exhibits the Klein paradox$^{18-19}$. However, no link between the Klein paradox and the negative refraction has been found in graphene.

\section{Left Handed Material}
To achieve our goal, we will first illustrate some relevant
features of the refraction at the interface between a positive refractive
index material (PIM) and effectively LHM. We will consider the Klein paradox later.

In Fig. 1, we show the negative refraction$^{1,14-16}$ of an obliquely incident Gaussian wavepacket, with a center frequency of $5{\rm GHz}$, at the interface between semi-infinite PIM ($z<0$) and NIM (i.e., LHM or negative index material) ($z>0$) media. From a simple SRR (common magnetic constitute of LHMs to provide the negative permeability) and wire pair-based LHM model$^{2,3}$, we estimated $\epsilon=-4.76$, $\mu=-1.222$, and $n=-2.412$ at the center frequency. Arrows indicate an arbitrary wavevector component, $\mathbf{K}_{L}$ ($L$ for LHM), in the incident wavepacket (see the bottom left quadrant of the figure) and the corresponding wavevector components, $\mathbf{Q}_{L}$ and $\mathbf{K'}_{L}$, in the transmitted (bottom right quadrant) and reflected (top left quandrant) wavepackets, respectively. Arrows are drawn perpendicular to their respective wavefronts (i.e., $\mathbf{K}_{L}$, for example, corresponds to a uniform plane wave component of the incident wavepacket in the bottom left quadrant). Below, we briefly describe why negative refraction occurs under the settings of Fig. 1.

Wave vector, $\mathbf{K}_{L}$, of the electric
field incident on the dispersive LHM from vacuum makes an
angle, $\Theta_{i}$, with the normal to the boundary located at $z=0$. Because the individual incident electric field is a plane
wave and linearly polarized along the $y$-axis, its wave vector
lies in the $xz$-plane. That is

\begin{equation}
E(\mathbf{r},t)=\textrm{e}^{i(\mathbf{K}_{L}\cdot\mathbf{r}-\omega t)},\end{equation}
with unit amplitude. It can be shown by the phase matching condition
and causality for the refracted wave that the transmitted wave vector $\mathbf{Q}_{L}$ is 

\begin{equation}
\mathbf{Q}_{L}={{K}_{L}}_{x}\hat{\mathbf{x}}+\sigma\sqrt{n^{2}{{K}_{L}}^{2}-{{{K}_{L}}_{x}}^{2}}\hat{\mathbf{z}}.\end{equation}
$\sigma=+1$ for the positive index medium and $-1$ for the LHM$^{14}$. From the definition of group velocity for isotropic,
low loss materials, the refracted beam has the group velocity

\begin{equation}
\mathbf{V}_{g}=\frac{\mathbf{Q}_{L}}{{Q}_{L}}\frac{\sigma c}{n_{g}},\end{equation}
where $c$ is the light speed and $n_{g}$ is the group index, which
is by causality always greater than unity$^{15}$. Thus the group
velocity is always antiparallel to the phase velocity of the refracted
wave for $\sigma=-1$. Since the time-averaged energy flux $\langle\mathbf{S}\rangle=\langle u\rangle\mathbf{V}_{g}$$^{16}$, where $\langle u\rangle$ is the time-averaged energy density,
it is clear from equations (2) and (3) and by causality that the incident
wave should undergo negative refraction.

Field transmission ($\tau_{L}$) and reflection ($\rho_{L}$) coefficients
for LHM can be determined by matching the electric and magnetic fields
at the interface between right and left handed media. That gives,

\begin{equation}
\tau_{L}=\frac{2\mu{{K}_{L}}_{z}}{\mu{{K}_{L}}_{z}+{{Q}_{L}}_{z}}\end{equation}

\begin{equation}
\rho_{L}=\frac{\mu{{K}_{L}}_{z}-{{Q}_{L}}_{z}}{\mu{{K}_{L}}_{z}+{{Q}_{L}}_{z}}.\end{equation}

\noindent where $\mu$ is the effective permeability of the LHM. Power transmission
and reflection coefficients can be obtained from the average energy
flux $\langle\mathbf{S}\rangle=(c/8\pi)\mathcal{\Re}[\mathbf{E\times}\mathbf{H}^{*}]$.
Thus we have, respectively,

\begin{equation}
T_{L}=\frac{|\tau_{L}|^{2}{{Q}_{L}}_{z}}{\mu{{K}_{L}}_{z}},\end{equation}

\begin{equation}
R_{L}=|\rho_{L}|^{2}.\end{equation}

\section{Klein Paradox}
Now, we turn our attention to the Klein paradox. Although the nonrelativistic
quantum mechanics of the scattering of a quantum particle is straightforward, the relativistic case displays quite peculiar situation
called the Klein Paradox$^{11,20-21}$. In the following we discuss 2D scattering
of a particle from a Klein step potential and consider some of these
peculiarities, which include transmission through strong potential
barrier, pair production, negative transmission and negative
refraction.

We consider a spin-$0$ particle with momentum $P$ and energy $E=(P^{2}c^{2}+m^{2}c^{4})^{1/2}$, which is governed by the KG equation,

\begin{equation}
[\mathbf{E}-V(\mathbf{r})]^{2}|\Psi\rangle=(c^{2}\mathbf{P}^{2}+m^{2}c^{4})|\Psi\rangle,\end{equation}
where $m$ is the mass of the particle. 

In Fig. 2, we show the negative refraction of an obliquely incident Gaussian beam of those spin-zero particles at the boundary of a strong potential, $V(\mathbf{r})$, which is assumed to be $0$ on the incident side and described by a 2D Klein step
potential, $V(\mathbf{r})=V$ for $z>0$. $\mathbf{K}_{K}$ represents the wave vector for an arbitrary plane wave component in the incident flux (see the bottom left quadrant of the figure), while $\mathbf{Q}_{K}$ and $\mathbf{K'}_{K}$ indicate the corresponding wave vectors for transmitted (bottom right quandrant) and reflected (top left quadrant) plane waves, respectively. The angle between $\mathbf{K}_{K}$ and the boundary normal is $\Theta_{i}$. As we demonstrate below, the directions of group ($\mathbf{V}_{g}$) and phase ($\mathbf{V}_{p}$) velocities for the transmitted beam are antiparallel.
 
In the position representation, equation (8) becomes

\begin{equation}
\{[i\hbar\partial_{t}-V(\mathbf{r})]^{2}+c^{2}\hbar^{2}\nabla^{2}-m^{2}c^{4}\}\Psi(\mathbf{r},t)=0.\end{equation}
For $z<0$, consider the individual positive energy plane wave component,  in the incident wavepacket in Fig. 2, which is of the form

\begin{equation}
\Phi(\mathbf{r},t)=\textrm{e}^{i(\mathbf{K}_{K}\cdot\mathbf{r}-Et/\hbar)}.\end{equation}
Note that equation (10) is similar to equation (1) given for the $y$-polarized
electric field.

We look for the solution of the form

\begin{equation}
\Psi(\mathbf{r})=\phi(-z)[\textrm{e}^{-iEt/\hbar}(\textrm{e}^{i\mathbf{K}_{K}\cdot\mathbf{r}}+\rho_{K}\textrm{e}^{i\mathbf{K}_{K}^{'}\cdot\mathbf{r}})]+\phi(z)\tau_{K}\textrm{e}^{-iEt/\hbar}\textrm{e}^{i\mathbf{Q}_{K}\cdot\mathbf{r}},\end{equation}
where $\phi(z)$ is the unit step function. Substitution of equation (11) into equation (9) and satisfying the phase matching condition, ${{K}_{K}}_{x}={{{K}^{'}}_{K}}_{x}={{Q}_{K}}_{x}$, yields

\begin{equation}
E^{2}=c^{2}\hbar^{2}{{K}_{K}}^{2}+m^{2}c^{4},\;\; z<0\end{equation}

\begin{equation}
c^{2}\hbar^{2}{{Q}_{K}}^{2}=(E-V)^{2}-m^{2}c^{4},\;\; z>0.\end{equation}
We can analyse equation (13) by considering three cases. If the potential
$V$ is weak, such that it is less than $E-mc^{2}$, then ${{Q}_{K}}$
has to be real. For the intermediate case, where $E-mc^{2}<V<E+mc^{2},$
${{Q}_{K}}$ is purely imaginary, so the trasmitted field is
damped.

We concentrate on the strong potential case, that is
$V>E+mc^{2}$. Note that we again have real ${{Q}_{K}}$. Even when
the energy of the incident particle is less than the height of the
potential, the transmitted particle doesn't necessarily undergo any
attenuation. This is a classically (also in nonrelativistic quantum mechanics) forbidden situation.

The question thus naturally arises: What is the direction of the momentum
of the transmitted wave, or $\mathbf{Q}_{K}$, for the strong potential
case? Since $E(\mathbf{Q}_{K})=E({{Q}_{K}})$ in equation (13),
for the group velocity of the transmitted wave we have

\begin{equation}
\mathbf{V}_{g}=\frac{\hbar c\mathbf{Q}_{K}}{E-V}.\end{equation}
If equation (14) is not considered for the moment, to satisfy the phase matching condition at
the interface, incoming wave in equation (10) should either propagate along the vector $\mathbf{Q}_{K}$ in Fig. 2 or along a vector, which is directed away from the boundary and has the same tangential component with that of the incoming ($\mathbf{K}_{K}$) and reflected waves ($\mathbf{K}_{K}^{'}$) (i.e., mirror symmetry of $\mathbf{Q}_{K}$ multiplied by $-1$). Since the denominator of equation (14) is negative, however, the group velocity $\mathbf{V}_{g}$ and the wave vector $\mathbf{Q}_{K}$ must be antiparallel. Therefore, we must pick the former path of propagation (i.e., along $\mathbf{Q}_{K}$). Otherwise, since the group velocity is the velocity of the moving wave packet, the causality is violated by allowing an incoming wavepacket from positive-$z$ side. Thus, it is clear that we observe negative refraction in Fig. 2 analogous to LHMs.

Since the momentum, or $\mathbf{Q}_{K}$, of the transmitted wave is
a function of the energy of the incident particle $E$, we can
write equation (13) in terms of its components as

\begin{equation}
c^{2}\hbar^{2}{{{Q}_{K}}_{z}}^{2}=c^{2}\hbar^{2}{{{K}_{K}}_{z}}^{2}-2EV+V^{2}.\end{equation}
Remember that we found ${{Q}_{K}}$ to be real in the strong
potential case, where $E<V$ and we know that ${{{Q}_{K}}_{x}}^{2}={{{K}_{K}}_{x}}^{2}$
is real positive. Therefore, we immediately conclude from equation (15) that ${{Q}_{K}}_{z}$
can be real as well as purely imaginary, even though ${Q}_{K}$
is real. Now we closely examine the condition which makes ${{Q}_{K}}_{z}$
either real or purely imaginary. To achieve that we define the excess
potential as

\begin{equation}
\Delta V\equiv V-E-mc^{2}.\end{equation}
If we substitute $V$ in equation (16), into equation (15) we obtain

\begin{equation}
c^{2}\hbar^{2}{{{Q}_{K}}_{z}}^{2}=2mc^{2}(\Delta V)+(\Delta V)^{2}-c^{2}\hbar^{2}{{{K}_{K}}_{x}}^{2}.\end{equation}

Since $\Delta V$ is arbitrarily chosen, for a given incident
particle with mass $m$, it is interesting that the stronger the potential
the more momentum transfer to the $z$-component of the transmitted
wave occurs without any damping. As we increase the height of the
potential step, keeping $m$ and ${{K}_{K}}_{x}$ the same,
the angle of refraction approaches zero. However, for the relatively
weaker potential such that $0<\Delta V<-mc^{2}+(c^{2}\hbar^{2}{{{K}_{K}}_{x}}^{2}+m^{2}c^{4})^{1/2}$,
the transmitted wave is damped. Interestingly this
does not occur in the 1D scattering from a strong potential of the
same kind. First, this could be easily understood from equation
(17) by setting ${{K}_{K}}_{x}=0$, so that ${{Q}_{K}}_{z}$
never becomes purely imaginary. Second, equation (17) also states that pair production in 2D scattering requires an additional potential
of $\Delta V=-mc^{2}+(c^{2}\hbar^{2}{{{K}_{K}}_{x}}^{2}+m^{2}c^{4})^{1/2}$
for given $m$ and ${{K}_{K}}_{x}$, compared to one-dimensional (1D) scattering.

The expressions for $\rho_{K}$ and $\tau_{K}$ in equation (11) are determined
by imposing the boundary conditions on $\Psi(\mathbf{r})$ and $\partial_{z}\Psi(\mathbf{r})$
at the interface, $z=0$. Thus we obtain

\begin{equation}
\tau_{K}=\frac{2{{K}_{K}}_{z}}{{{K}_{K}}_{z}+{{Q}_{K}}_{z}}\end{equation}

\begin{equation}
\rho_{K}=\frac{{{K}_{K}}_{z}-{{Q}_{K}}_{z}}{{{K}_{K}}_{z}+{{Q}_{K}}_{z}}.\end{equation}
We find the transmission $T_{K}$ and reflection $R_{K}$ coefficients
from the probability current, which is given by

\begin{equation}
\mathbf{J}=\frac{1}{2im}(\Psi^{\star}\nabla\Psi-\Psi\nabla\Psi^{\star}).\end{equation}
If ${{Q}_{K}}_{z}$ is purely imaginary (i.e., $relatively$
$weaker$ strong potential), we get

\begin{equation}
R_{K}=|\rho_{K}|^{2}=1,\;\; T_{K}=0.\end{equation}
This is the same result with the intermediate Klein step in 1D. If
${{Q}_{K}}_{z}$ is real, however, the coefficients are

\begin{equation}
T_{K}=\frac{{{Q}_{K}}_{z}|\tau_{K}|^{2}}{{{K}_{K}}_{z}}\end{equation}

\begin{equation}
R_{K}=|\rho_{K}|^{2}.\end{equation}
Note, since ${{Q}_{K}}_{z}<0$, the probability is conserved at the expense of a negative
transmission coefficient and a reflection coefficient exceeding unity.
This paradox can be resolved by employing the notion of particle-antiparticle
pair production, due to the strong potential. The antiparticles create
a negative charged current moving right inside the potential barrier,
while the particles are reflected and combined with the incident beam
leading to a positively charged current moving to the left. This is
the essence of the peculiarities in equations (22) and (23).

\section{Conclusion}
We theoretically investigated the underlying physics of similar behaviors resulting from Klein-Gordon and Maxwell's equations in the negative refraction regime. To the best of our knowledge, such a 2D scattering from a semi-infinite Klein step potential is treated here first, inspired by LHMs. The 2D case not only verifies the Klein paradox existing in one dimension, but also manifests the negative refraction phenomenon for matter waves. Based on the analogy described in this letter, we simulated the Klein paradox using LHMs. In fact, although not mentioned above, Fig. 2 is retrieved directly from the LHM simulation in Fig. 1 under certain transformations. Basically, individual Fourier components of the wavepackets in both the LHM and Klein cases are analytically matched. This directly maps the refractive index of the LHM to the Klein potential. A constant Klein potential given by the transformation, despite the inherent dispersion in the LHM, is ensured by the counter-dispersive nature of the mapping that compensates the LHM dispersion. Our work has implications to the analog quantum simulations of many exotic phenomena in the field of relativistic quantum mechanics with relatively simple optical benchtop experiments, incorporating an appropriate metamaterial design or graphene.

\section*{Acknowledgments}
Authors would like to thank David R. Smith and Thomas Koschny for inspiring and fruitful discussions. This work was supported in part by the National Security 
Agency (NSA) and Advanced Research and Development Activity (ARDA) 
under Army Research Office (ARO) Grant No.\ DAAD19-01-1-0520, by 
the DARPA QuIST program under contract F49620-02-C-0010, and by
the National Science Foundation (NSF) under grant ECS-0202087.

\newpage
\section*{List of Figure Captions}

Fig. 1. (Color online) Negative refraction of an obliquely incident Gaussian wavepacket at the interface between semi-infinite PIM and NIM media. Arrows indicate the wavevectors for an arbitrary plane wave ($\mathbf{K}_{L}$) in the incident wavepacket and its corresponding transmitted ($\mathbf{Q}_{L}$) and reflected ($\mathbf{K'}_{L}$) wavevectors. They are drawn perpendicular to their respective wavefronts. The colors show the relative intensity pattern at an arbitrary time. Lengths are in SI units. Interference fringes occur due to the incident (bottom left) and reflected (top left) wavepackets near the interface. Transmittance and reflectance at the center frequency, $\omega=5{\rm GHz}$, gives $T_{L}=0.855$ and $R_{L}=0.145$, respectively, for $\Theta_{i}=\pi/6$.

\noindent Fig. 2. (Color online) Negative refraction of a Gaussian beam of spin-zero particles at the boundary of a strong potential, $V$, for $z>0$. $\mathbf{K}_{K}$ indicates the wave vector of an arbitrary wave component in the incident flux, while $\mathbf{Q}_{K}$ and $\mathbf{K'}_{K}$ are the corresponding transmitted and reflected  counterparts. The directions of group ($\mathbf{V}_{g}$) and phase ($\mathbf{V}_{p}$) velocities are antiparallel. The colors show the relative probability density pattern at an arbitrary time. Lengths are in SI units. Transmittance and reflectance are calculated as $T_{K}=-3.66$ and $R_{K}=4.66$, respectively, at the center energy, $E=20.7 {\rm \mu}eV$, of the incoming wavepacket, taking $V=70.63{\rm \mu}eV$ and $\Theta_{i}=\pi/6$.

\newpage
\begin{figure}[tbph]
\centering
\includegraphics[%
  scale=0.60]{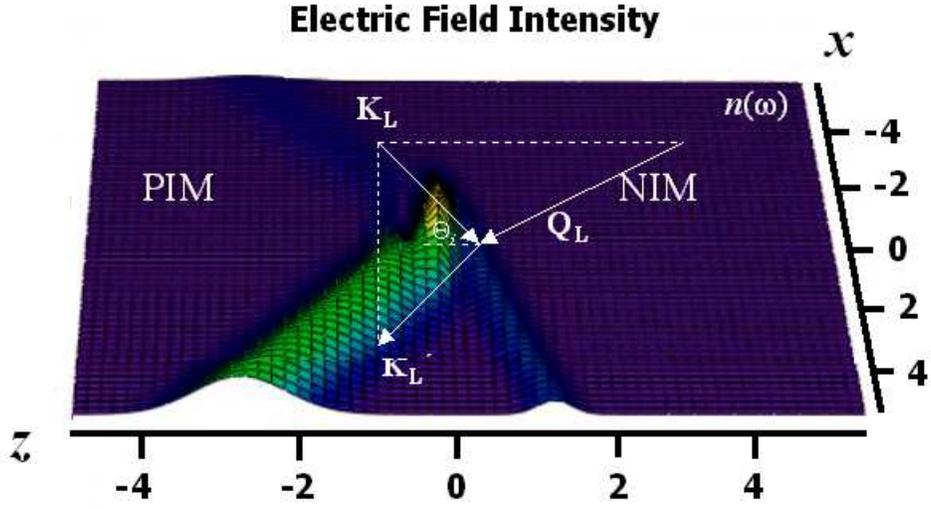}

\caption{(Color online) Negative refraction of an obliquely incident Gaussian wavepacket at the interface between semi-infinite PIM and NIM media. Arrows indicate the wavevectors for an arbitrary plane wave ($\mathbf{K}_{L}$) in the incident wavepacket and its corresponding transmitted ($\mathbf{Q}_{L}$) and reflected ($\mathbf{K'}_{L}$) wavevectors. They are drawn perpendicular to their respective wavefronts. The colors show the relative intensity pattern at an arbitrary time. Lengths are in SI units. Interference fringes occur due to the incident (bottom left) and reflected (top left) wavepackets near the interface. Transmittance and reflectance at the center frequency, $\omega=5{\rm GHz}$, gives $T_{L}=0.855$ and $R_{L}=0.145$, respectively, for $\Theta_{i}=\pi/6$.}

\end{figure}

\newpage
\begin{figure}[tbph]
\centering
\includegraphics[%
  scale=0.60]{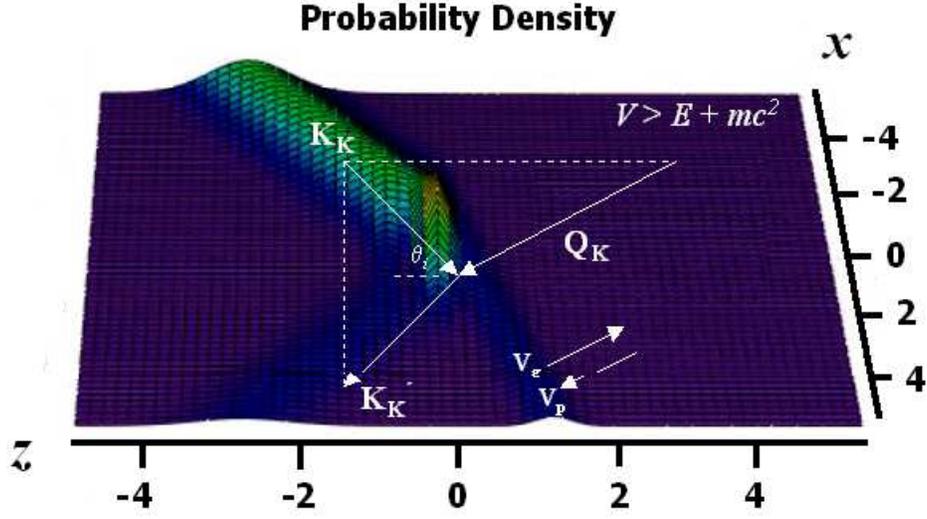}

\caption{(Color online) Negative refraction of a Gaussian beam of spin-zero particles at the boundary of a strong potential, $V$, for $z>0$. $\mathbf{K}_{K}$ indicates the wave vector of an arbitrary wave component in the incident flux, while $\mathbf{Q}_{K}$ and $\mathbf{K'}_{K}$ are the corresponding transmitted and reflected  counterparts. The directions of group ($\mathbf{V}_{g}$) and phase ($\mathbf{V}_{p}$) velocities are antiparallel. The colors show the relative probability density pattern at an arbitrary time. Lengths are in SI units. Transmittance and reflectance are calculated as $T_{K}=-3.66$ and $R_{K}=4.66$, respectively, at the center energy, $E=20.7 {\rm \mu}eV$, of the incoming wavepacket, taking $V=70.63{\rm \mu}eV$ and $\Theta_{i}=\pi/6$.}

\end{figure}

\end{document}